\documentclass[sigconf,authorversion,nonacm]{acmart}

\usepackage{amsmath,amsfonts,bm}
\usepackage{algorithmic}
\usepackage[noend,figure]{algorithm2e}
\usepackage{setspace}
\usepackage{graphicx}
\usepackage{tikz}
\usepackage{multirow}
\usepackage{textcomp}
\usepackage{capt-of}
\usepackage{xcolor}
\usepackage{subfig}
\usepackage{colortbl}
\usepackage{rotating}
\usepackage{booktabs}
\usepackage{multirow}
\usepackage{mathtools}
\usepackage{siunitx}
\usepackage{array}
\usepackage{numprint}
\usepackage{xurl}

\def\HiLi{\leavevmode\rlap{\hbox to \hsize{\color{gray!15}\leaders\hrule height .7\baselineskip depth .5ex\hfill}}}
\def\HiLiS{\leavevmode\rlap{\hbox to 0.05\hsize{\color{gray!15}\leaders\hrule height .7\baselineskip depth .5ex\hfill}}}

\AtBeginDocument{%
  }

\usepackage{xcolor}
\usepackage{listings}

\definecolor{codegreen}{rgb}{0,0.8,0}
\definecolor{codegray}{rgb}{0.5,0.5,0.5}
\definecolor{codepurple}{rgb}{0.58,0,0.82}
\definecolor{backcolour}{rgb}{0.95,0.95,0.92}
\definecolor{grayout}{rgb}{0.5,0.5,0.5}
\definecolor{colgray}{gray}{0.93}

\newcommand{\mysb}{\textsc{SparseBoomerang}}
\newcommand{\myb}{\textsc{Boomerang}}
\newcommand{\myfd}{\textsc{FlowDroid}}

\newcommand{\mysh}{\textsc{SparseHeros}}
\newcommand{\myh}{\textsc{Heros}}
\newcommand{\mybench}{\textsc{ConstantBench}}
\newcommand{\mysoot}{\textsc{Soot}}
\newcommand{\myjimple}{\textsc{Jimple}}

\newcommand{\rqone}{RQ1: Does Sparse IDE produce the same results as the original IDE?}
\newcommand{\rqtwo}{RQ2: How does the sparsification impact the performance in terms of runtime and memory?}
\newcommand{\rqthree}{RQ3: To what extent does the number of propagations correlate with the performance impact?}

\newcommand{\rom}[1]{\uppercase\expandafter{\romannumeral #1\relax}}

\lstdefinestyle{mystyle}{
    commentstyle=\color{codegreen},
    keywordstyle=\color{black},
    numberstyle=\tiny\color{codegray},
    stringstyle=\color{codepurple},
    basicstyle=\ttfamily\footnotesize,
    breakatwhitespace=false,         
    breaklines=true,
    lineskip=0.5cm,
    captionpos=b,                    
    keepspaces=true,                 
    numbers=left,                    
    numbersep=5pt,                  
    showspaces=false,                
    showstringspaces=false,
    showtabs=false,                  
    tabsize=2
}

\newcommand{\mycomment}[1]{}

\begin{document}

\title{Symbol-Specific Sparsification\\ of Interprocedural Distributive Environment Problems}

\author{Kadiray Karakaya}
\orcid{0000-0001-9266-2084}
\affiliation{%
  \institution{Heinz Nixdorf Institute\\ Paderborn University}
  \city{Paderborn}
  \country{Germany}
}
\email{kadiray.karakaya@upb.de}

\author{Eric Bodden}
\orcid{0000-0003-3470-3647}
\affiliation{%
  \institution{Heinz Nixdorf Institute\\ Paderborn University \& Fraunhofer IEM}
  \city{Paderborn}
  \country{Germany}}
\email{eric.bodden@upb.de}

\renewcommand{\shortauthors}{Karakaya and Bodden}

\begin{abstract}
    Previous work has shown that one can often greatly speed up static analysis by computing data flows not for every edge in the program's control-flow graph but instead only along definition-use chains. This yields a so-called \emph{sparse} static analysis. Recent work on \textsc{SparseDroid} has shown that specifically taint analysis can be ``sparsified'' with extraordinary effectiveness because the taint state of one variable does not depend on those of others. This allows one to soundly omit more flow-function computations than in the general case.
    
    In this work, we now assess whether this result carries over to the more generic setting of so-called Interprocedural Distributive Environment (IDE) problems. Opposed to taint analysis, IDE comprises distributive problems with large or even \emph{infinitely broad} domains, such as typestate analysis or linear constant propagation. 
    Specifically, this paper presents
    Sparse IDE, a framework that realizes sparsification for any static analysis that fits the IDE framework.
   
    We implement Sparse IDE in \mysh{}, as an extension to the popular \myh{} IDE solver, and evaluate its performance on real-world Java libraries
    by comparing it to the baseline IDE algorithm.
    To this end, we design, implement and evaluate a linear constant propagation analysis client on top of \mysh{}.
    Our experiments show that, although IDE analyses can only be sparsified with respect to symbols and not (numeric) values, Sparse IDE can nonetheless yield significantly lower runtimes and often also memory consumptions compared to the original IDE.    
\end{abstract}

\keywords{static analysis, sparse analysis, IFDS, IDE, constant propagation}

\maketitle

\section{Introduction}
Static program analysis has proven useful for diverse purposes including compiler optimization \cite{optimization}, program comprehension \cite{compreh} and developer assistance \cite{assist}. 
It is now an essential part of software engineering for assuring bug-free \cite{bugs}, secure \cite{vuln2} and quality software\cite{qual}. 
The key strength of static program analysis is to account for all possible executions of a target program. 
But this imposes two often competing challenges: precision and scalability.
Static analyses yield more precise results by tracking statement ordering and by distinguishing different calling contexts.

IDE (Interprocedural Distributive Environment) \cite{ide}, with its extensions~\cite{extendIFDS, reviser, ideal},
is a state-of-the-art precise interprocedural static analysis framework.
It covers a wide class of data-flow problems ranging from variations of classical taint analysis \cite{vuln} to typestate~\cite{tsf, li2022path} and constant propagation \cite{icc} analyses. IDE represents data-flow analysis problems on an \emph{exploded supergraph} and models data-flow facts as environments. Environments are mappings from symbols (often program variables) to domain values. The exploded supergraph is a data-flow graph induced by the inter-procedural control-flow graph (ICFG) for the whole program. Its nodes are pairs $(s,d)$ of program statements and data-flow facts. 
A data-flow fact $d$ holds at a statement $s$ if in the exploded supergraph the corresponding node $(s,d)$ is reachable from the start node. The edges of the exploded supergraph represent the effects of program statements on a data-flow fact. IDE computes over the exploded supergraph by tracking all data-flow facts \emph{densely} across all program points. As previous work~\cite{sparsedroid,cleandroid,diskdroid,gpudroid} has shown, this approach does not scale well for large-scale real-world programs. A key observation is, however, that in practice many program statements do not affect the analysis result. Such statements thus can be safely ignored, e.g.\ by \emph{sparsifying} the exploded supergraph. 

Sparsification is a well-known technique for scaling data-flow analyses \cite{pinpoint,svf,hardekopf2011millions,sparse-global,spas,semi-sparse} while still maintaining their precision. Sparsification approaches create sparse versions of the original CFGs of a target program by removing statements that are irrelevant to the analysis and then computing over the sparse CFGs.
Recent on-demand approaches take sparsification further by utilizing the information available during the analysis. 
\mysb{}~\cite{sparseBoomerang} accelerates demand-driven pointer analysis by computing over sparse CFGs specialized to the alias queries. 
\textsc{SparseDroid} \cite{sparsedroid} accelerates taint analysis by computing over sparse CFGs specialized to individual data-flow facts. Both approaches demonstrate sparsification on IFDS-based problems, that focus on mere symbol reachability, without considering value computation.

The IFDS (Interprocedural Finite Distributive Subset) \cite{ifds} framework is the ``small brother'' of IDE. It reduces the data-flow analysis problems to a pure graph reachability problem. 
Yet, IFDS is limited to data-flow problems with finite domains: all IFDS problems can be encoded as IDE problems, but only a subset of IDE problems can be encoded as IFDS problems~\cite{ide}. 
As an example, consider the statement \texttt{a = a + 1}. Here, using IFDS one can encode a simple taint analysis inferring that \texttt{a} is tainted/reachable after the statement if and only if it was previously tainted/reachable. Efficient computation of \texttt{a}'s numeric value, however, requires one to \emph{compute values} within the infinitely broad domain of integers, going beyond pure reachability. 
As we show, this has implications for sparsification: while the statement \texttt{a = a + 1} can be safely considered irrelevant w.r.t.\ \texttt{a}'s reachability, 
and will be disregarded in sparsification approaches for IFDS~\cite{sparseBoomerang,sparsedroid}, 
it \emph{is} a relevant statement when constant propagation is considered: it changes \texttt{a}'s value. 
This observation is not limited to constant propagation analysis, it applies to other data-flow analysis problems that require value mappings. 
For instance, a sparse typestate analysis must retain statements that alter a symbol's associated state value. 
Based on this observation, we generalize the recent work on \textsc{SparseDroid}, i.e., on sparse IFDS~\cite{sparsedroid}: 
we propose \emph{Sparse IDE}, a symbol-specific sparsification of the IDE framework, that enables efficient sparsification, even in the presence of arbitrarily large value domains.
In addition, we also show the limits of sparsification in IDE: while one can effectively sparsify with respect to symbols, such sparsification cannot be performed with respect to values.

We formalize Sparse IDE, and show how this formalization covers also IFDS data-flow analysis problems as a special case.
We implement Sparse IDE in a tool \textsc{SparseHeros}, extending the popular \textsc{Heros} IDE solver~\cite{heros}. 
We compare both implementations in terms of performance, and show that sparsification maintains correctness.
To this end, we implement a linear constant propagation analysis client that uses both implementations. To validate \textsc{SparseHeros}'s correctness, we run both on \mybench{}, a novel microbenchmark suite for integer linear constant propagation analysis.
To evaluate its performance impact, we run the analysis client on real-world Java libraries using both \myh{} and \mysh{}.
The analysis client produces the same results in both cases while terminating significantly faster when using \mysh{}.

To summarize, this paper presents the following original contributions, whose implementations are open-sourced\footnote{https://github.com/secure-software-engineering/SparseIDE}:
\begin{itemize}
    \item A formalization of Sparse IDE  and its implementation in \mysh{} on top of \myh{} and \mysoot{}~\cite{soot},
    \item its correctness evaluation on the \mybench{} microbenchmark suite for linear constant propagation analysis, and
    \item its performance evaluation on real-world Java libraries.
\end{itemize}

The remainder of the paper is organized as follows. In Section \ref{sec:bg}, we present the background. 
In Section \ref{sec:approach}, we introduce Sparse IDE and in Section \ref{sec:impl}, 
we instantiate it on linear constant propagation analysis. 
In Section \ref{sec:eval}, we present the evaluation results. 
In Section \ref{sec:threats}, we discuss the limitations of our approach and threats to its validity. 
In Section \ref{sec:related}, we discuss the related work and we conclude with Section \ref{sec:conc}.

\section{Background} \label{sec:bg}
This section briefly introduces the background that our work builds on. We begin with the IFDS and IDE frameworks. Then we introduce sparse data-flow analysis and discuss why it is an effective alternative. Finally, we explain how the recent approaches sparsify further by utilizing the information available during the analysis runtime.

\begin{table}[h!]
    \centering
    \resizebox{0.9\columnwidth}{!}{%
    \begin{tabular}{|c|c|c|} 
    \hline
    \(f_{id}\): \(\lambda S.S\) & \(f_{gen}\): \(\lambda S.(S \cup \{a\})\) & \(f_{as}\): \(\lambda S\).if \(a \in S\): \((S \cup \{b\})\) else  \((S \setminus \{b\}) \)  \\
    \hline
\begin{tikzpicture}
    \node[circle,fill,inner sep=1.2pt, label={above:\(\Lambda\)}] (z1) at (0,0) {};
    \node[circle,fill,inner sep=1.2pt, label={[label distance=0.5pt]below:\(\Lambda\)}] (z2) at (0,-1){};
    \draw[->,shorten <=1pt, shorten >=1pt] (z1) -- (z2);

    \node[circle,fill,inner sep=1.2pt, label={above:\emph{a}}] (a1) at (0.6,0) {};
    \node[circle,fill,inner sep=1.2pt, label={[label distance=1.8pt]below:\emph{a}}] (a2) at (0.6,-1){};
    \draw[->,shorten <=1pt, shorten >=1pt] (a1) -- (a2);
\end{tikzpicture} 
& 
\begin{tikzpicture}
    \node[circle,fill,inner sep=1.2pt, label={above:\(\Lambda\)}] (z1) at (0,0) {};
    \node[circle,fill,inner sep=1.2pt, label={[label distance=0.5pt]below:\(\Lambda\)}] (z2) at (0,-1){};
    \draw[->,shorten <=1pt, shorten >=1pt] (z1) -- (z2);

    \node[circle,fill,inner sep=1.2pt, label={above:\emph{a}}] (a1) at (0.6,0) {};
    \node[circle,fill,inner sep=1.2pt, label={[label distance=1.8pt]below:\emph{a}}] (a2) at (0.6,-1){};
    \draw[->,shorten <=1pt, shorten >=1pt] (z1) -- (a2);
\end{tikzpicture} 
& 
\begin{tikzpicture}
    \node[circle,fill,inner sep=1.2pt, label={above:\(\Lambda\)}] (z1) at (0,0) {};
    \node[circle,fill,inner sep=1.2pt, label={[label distance=0.5pt]below:\(\Lambda\)}] (z2) at (0,-1){};
    \draw[->,shorten <=1pt, shorten >=1pt] (z1) -- (z2);

    \node[circle,fill,inner sep=1.2pt, label={above:\emph{a}}] (a1) at (0.6,0) {};
    \node[circle,fill,inner sep=1.2pt, label={[label distance=1.8pt]below:\emph{a}}] (a2) at (0.6,-1){};
    \draw[->,shorten <=1pt, shorten >=1pt] (a1) -- (a2);

    \node[circle,fill,inner sep=1.2pt, label={above:\emph{b}}] (b1) at (1.2,0) {};
    \node[circle,fill,inner sep=1.2pt, label={below:\emph{b}}] (b2) at (1.2,-1){};
    \draw[->,shorten <=1pt, shorten >=1pt] (a1) -- (b2);
\end{tikzpicture} 
\\
    \hline
    \end{tabular}
    }

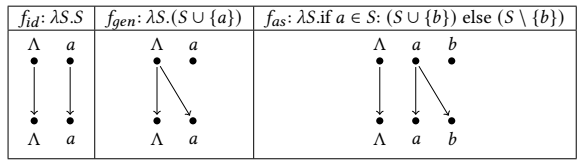
\captionof{figure}{Flow functions (reproduced from \cite{ifds}).}\label{fig_ff}
\end{table}

\subsection{IFDS and IDE}
IFDS \cite{ifds} and IDE \cite{ide} are two frameworks for interprocedural flow- and context-sensitive data-flow analysis.
IFDS represents data-flow analysis problems as graph reachability on an exploded supergraph, whose nodes are pairs of program statements and data-flow facts. 
The individual edges in the exploded supergraph constitute \emph{flow functions}; they show each statement's effect on each data-flow fact's reachability. A flow function determines whether a data-flow fact is being generated, propagates to the next statement, spawns another fact, or gets killed. 

Figure \ref{fig_ff} shows how the flow functions are represented as edges in the exploded supergraph.
The data-flow fact above the edge means that it holds before applying the function; the fact below means that it holds after. A special fact, \(\Lambda\) holds always. Facts connected to it are newly generated. 
The identity function, \(f_{id}\), leaves data-flow facts unchanged. The function \(f_{gen}\) shows the case where data-flow fact \emph{a} is being generated.
The function \(f_{as}\) shows how the existing fact, \emph{a} creates another fact, \emph{b}, e.g. at an assignment, \texttt{b = a}.

IDE generalizes the IFDS framework by computing domain values that symbols map to. It does so in two phases: first it determines whether symbols are reachable, just like IFDS, and then computes their values. IDE achieves this by annotating the individual exploded supergraph edges with so-called \emph{edge functions}, which constitute environment transformers. 

\begin{table}[h!]
    \centering
    \resizebox{0.9\columnwidth}{!}{%
    \begin{tabular}{|c|c|c|} 
    \hline
    \(e_{id}\): \(\lambda env.env\) & \(e_{val}\): \(\lambda env.env[a \mapsto 3]\) & \(e_{op}\): \(\lambda env.env[b \mapsto 2*env(a) + 1]\)  \\
    \hline
\begin{tikzpicture}
    \node[circle,fill,inner sep=1.2pt, label={above:\(\Lambda\)}] (z1) at (0,0) {};
    \node[circle,fill,inner sep=1.2pt, label={[label distance=0.5pt]below:\(\Lambda\)}] (z2) at (0,-1){};
    \draw[->,shorten <=1pt, shorten >=1pt] (z1) -- node[midway,right,pos=0.7,xshift=-3pt] {\(\lambda l.l\)} (z2);

    \node[circle,fill,inner sep=1.2pt, label={above:\emph{a}}] (a1) at (0.8,0) {};
    \node[circle,fill,inner sep=1.2pt, label={[label distance=1.8pt]below:\emph{a}}] (a2) at (0.8,-1){};
    \draw[->,shorten <=1pt, shorten >=1pt] (a1) -- node[midway,right,pos=0.7,xshift=-3pt] {\(\lambda l.l\)} (a2);
\end{tikzpicture} 
& 
\begin{tikzpicture}
    \node[circle,fill,inner sep=1.2pt, label={above:\(\Lambda\)}] (z1) at (0,0) {};
    \node[circle,fill,inner sep=1.2pt, label={[label distance=0.5pt]below:\(\Lambda\)}] (z2) at (0,-1){};
    \draw[->,shorten <=1pt, shorten >=1pt] (z1) -- node[midway,right,pos=0.7,xshift=-3pt] {\(\lambda l.l\)} (z2);

    \node[circle,fill,inner sep=1.2pt, label={above:\emph{a}}] (a1) at (1,0) {};
    \node[circle,fill,inner sep=1.2pt, label={[label distance=1.8pt]below:\emph{a}}] (a2) at (1,-1){};
    \draw[->,shorten <=1pt, shorten >=1pt] (z1) -- node[midway,right,pos=0.7,xshift=-2pt] {\(\lambda l.3\)} (a2);
\end{tikzpicture} 
& 
\begin{tikzpicture}
    \node[circle,fill,inner sep=1.2pt, label={above:\(\Lambda\)}] (z1) at (0,0) {};
    \node[circle,fill,inner sep=1.2pt, label={[label distance=0.5pt]below:\(\Lambda\)}] (z2) at (0,-1){};
    \draw[->,shorten <=1pt, shorten >=1pt] (z1) -- node[midway,right,pos=0.7,xshift=-3pt] {\(\lambda l.l\)} (z2);

    \node[circle,fill,inner sep=1.2pt, label={above:\emph{a}}] (a1) at (0.8,0) {};
    \node[circle,fill,inner sep=1.2pt, label={[label distance=1.8pt]below:\emph{a}}] (a2) at (0.8,-1){};
    \draw[->,shorten <=1pt, shorten >=1pt] (a1) -- node[midway,right,pos=0.7,xshift=-3pt] {\(\lambda l.l\)} (a2);

    \node[circle,fill,inner sep=1.2pt, label={above:\emph{b}}] (b1) at (1.8,0) {};
    \node[circle,fill,inner sep=1.2pt, label={below:\emph{b}}] (b2) at (1.8,-1){};
    \draw[->,shorten <=1pt, shorten >=1pt] (a1) -- node[midway,right,pos=0.7,xshift=1pt] {\(\lambda l.2*l+1\)} (b2);
\end{tikzpicture} 
 \\
    \hline
    \end{tabular}
    }

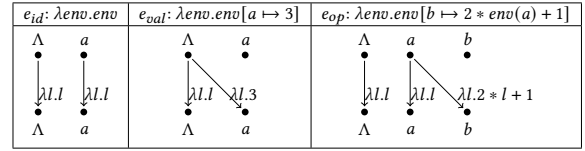
\captionof{figure}{Edge functions (reproduced from \cite{ide}).}\label{fig_ef}
\end{table}

Figure \ref{fig_ef} shows how the edge functions are represented. The environment transformer \(e_{id}\) keeps the values as they are. \(e_{val}\) shows the case where data-flow fact, \emph{a} is mapped to a domain value, e.g. through a constant assignment, \texttt{a = 3}. \(e_{op}\) shows how the value of \emph{b} is calculated depending on the value of \emph{a}, e.g. through a linear arithmetic operation, \texttt{b = 2*a + 1}. IDE can only compute \emph{linear} equations precisely.

IFDS and IDE apply to a wide class of data-flow analysis problems.
IFDS requires data-flow problems to be defined with flow functions that are distributive over the merge operator. Many reachability problems such as taint, reaching definitions, or live variables analysis fall into this category.
IDE, on the other hand, also requires data-flow problems to be expressed with distributive environment transformers. 
IFDS suits better the problems with a binary value domain, e.g. taint analysis where the domain simply consists of two values, \emph{tainted} or \emph{not tainted} \cite{flowdroid}. It has been applied to more complex domains, e.g. for typestate analysis where the domain contains arbitrary object states \cite{typestate-ifds}. The drawback of IFDS is that it represents data-flow facts as symbol-value pairs, which blows up the data-flow fact space with increasing size of the domain. Because of this representation, IFDS's runtime performance depends on the value domain's size. Further, it may not terminate when the value domain is infinitely broad, e.g., in constant propagation analysis, where the domain contains all integers. IDE, on the other hand, restricts data-flow facts to static symbols and computes their (approximated) runtime values using the edge functions along the path where the symbols are reachable in the exploded supergraph. Therefore, IDE can terminate efficiently even with infinitely broad value domains---only the set of symbols must be finite.

\subsection{Sparse Data-flow Analysis}
Data-flow analysis techniques aim to produce precise results while remaining scalable within a reasonable time budget. Techniques that prioritize scalability often resort to sacrificing precision aspects: flow-insensitive analyses ignore control-flow ordering \cite{selective_flow}, field-insensitive analyses approximate field accesses\cite{access-graphs}, and context-insensitive analyses do not distinguish different calling contexts \cite{selective_context}. Sparse data-flow analyses, on the other hand, often improve a \emph{dense} data-flow analysis' scalability while \emph{maintaining} its precision. They sparsify a target program's control-flow graph by removing program statements that provably do not affect the analysis result. Sparsification often uses a cheaper pre-analysis stage to aid a more expensive analysis~\cite{pinpoint, svf, hardekopf2011millions}. Recent \emph{on-demand} sparse data-flow analyses sparsify further by exploiting the information that is only available during analysis runtime ~\cite{sparseBoomerang, sparsedroid}.

\subsection{Fact-Specific On-Demand Sparsification}
\begin{figure}[h!]
    \centering
    \includegraphics[width=0.8\linewidth,keepaspectratio]{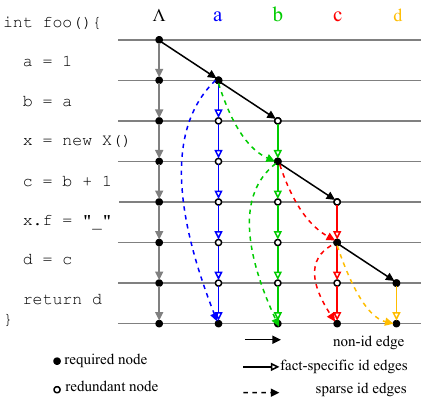}
    \captionsetup{skip=5pt,belowskip=-10pt}
    \caption{Original and sparse propagations after applying fact-specific on-demand sparsification.}
    \label{fig:sparseifds}
\end{figure}

When IFDS and IDE compute a data-flow fact's reachability, starting from the statement that generates the data-flow fact, they propagate it along all statements as long as it is not killed. At each statement, they check whether the statement is relevant for all the data-flow facts that have reached it. Figure~\ref{fig:sparseifds} shows how the reachability is computed for an example constant-propagation analysis setting. The \emph{fact-specific id edges} and \emph{non-id edges} show the edges which IFDS and IDE create when propagating data-flow facts. The data-flow facts actually only need to be propagated to the  \emph{required nodes}. For instance, data-flow fact \textbf{a} only needs to propagate to the statement \texttt{b = a;} all other statements are redundant for \textbf{a}. Similarly, \textbf{b} only needs to propagate to the statement, \texttt{c = b + 1}. Based on this observation, He et al. \cite{sparsedroid} introduced the sparse IFDS algorithm in their implementation \textsc{SparseDroid}. Instead of propagating all the data-flow facts to the next statement, it propagates them simply to the next statement that uses the facts. Sparse IFDS keeps all \emph{non-id edges} and replaces the \emph{fact-specific id edges} with \emph{sparse id edges}, effectively keeping all \emph{required nodes} and skipping over all \emph{redundant nodes}.

Fact-specific on-demand sparsification allows effective propagation of the data-flow facts along the sparse CFGs specific to them, which is not limited to data-flow analysis. Recent work \cite{sparseBoomerang} has applied it to pointer analysis, where the variable in alias queries is treated as the initial data-flow fact and propagated along its query-specific sparse CFGs. So far, however, fact-specific on-demand sparsification has only been applied to the analysis problems that deal with fact reachability. In this work, we expand the scope of fact-specific on-demand sparsification to include the data-flow analyses that compute over an additional value domain, specifically IDE.

\section{Symbol-specific On-Demand Sparsificiation with Sparse IDE} \label{sec:approach}
In this section, we first explain the original IDE algorithm \cite{ide} in detail. 
We then introduce the Sparse IDE algorithm by highlighting the modifications to the original IDE algorithm. 

\subsection{The Original IDE Algorithm}
Sagiv et al.~\cite{ide} define an IDE problem instance formally as $IP = (G^*,D,L,M)$, where 
\begin{itemize}
    \item $G^*$ is the program supergraph (ICFG), 
    which consists of control flow graphs (CFG), $G_p$ of individual procedures, 
    \item $D$ is a finite set of program symbols,
    \item $L$ is a finite-height lattice (which can be infinitely broad), and
    \item $M~:~E^* \xrightarrow[]{d} (Env(D,~L) \rightarrow Env(D,~L)) $ is an assignment of distributive environment transformers to the edges of $G^*$.
\end{itemize}

The original IDE algorithm \cite{ide} solves such an IDE problem, $IP$, in two phases.
In Phase \rom{1}, it creates the jump functions that show the reachability of each $d \in D$, 
by assuming that their initial mappings to $L$ are always $\lambda l.\top$.
In Phase \rom{2}, it computes each $d$'s actual value mapping to $L$ by evaluating the edge functions defined in $M$.

According to Sagiv et al. \cite{ide}, the total cost of the IDE algorithm is bounded by $O(|E||D|^3)$, which is the cost of Phase~\rom{1}. Since $D$ is the set of symbols, it should not change if correctness is preserved. We, therefore, apply our sparsification approach in Phase \rom{1}, where the jump functions are created by reducing $E$, the set of edges. Phase~\rom{2} is oblivious to how the jump functions are created---it automatically benefits from the sparsification of Phase~\rom{1}.

Figure \ref{ide-p1} shows the algorithm for Phase \rom{1}.
Each procedure $p$'s CFG, $G_p$ consists of a \emph{start} node $s_p$, an \emph{exit} node $e_p$, and \emph{normal} (non-call) nodes $m$ or $n$. Procedure calls are represented with two nodes: the \emph{call-site} node $c$ denotes the point right before the procedure call, and the \emph{return-site} node $r$ denotes the point right after.
Program symbols, e.g.\ variables, access paths, etc., are denoted with $d', d \in D \cup \{\Lambda\}$ including the special symbol $\Lambda$. $\Lambda$ is required for generating new symbols at arbitrary program points.

\LinesNumbered
\DontPrintSemicolon
\begin{algorithm}[t]
    \setstretch{1.35}
    \SetInd{0em}{0.5em}
    \scriptsize
    \SetKwFunction{FMain}{ForwardComputeJumpFunctionsSLRPs}
    \SetKwFunction{FPropagate}{Propagate}
    \SetKwData{Variable}{variable}
    \SetKwData{Value}{value}
    \SetKwProg{Fn}{Function}{:}{}
    \SetKwProg{Pn}{Function}{:}{}
    \Pn{\FMain{}}{
        \For{
            $\langle s_p, d'\rangle$, $\langle m, d\rangle$ s.t. $m$ occurs in proc. $p$ and $d', d \in D 	\cup \{\Lambda\}$
        }{
            $JumpFn(\langle s_p, d'\rangle \rightarrow \langle m, d\rangle) = \lambda l.\top$
        }
        \For{
            corresponding call-return pairs $(c,r)$ and $d', d \in D 	\cup \{\Lambda\}$
        }{
            $SummaryFn(\langle c, d'\rangle \rightarrow \langle r, d\rangle) = \lambda l.\top$
        }
        \emph{PathWorkList} $ \coloneqq \{\langle s_{main}, \Lambda\rangle\ \rightarrow \langle s_{main}, \Lambda\rangle \}$ \\
        $JumpFn(\langle s_{main}, \Lambda\rangle\ \rightarrow \langle s_{main}, \Lambda\rangle ) \coloneqq id$ \\
        \While{
            PathWorkList $ \neq \emptyset $
        }{
            Select and remove an item $\langle s_p, d_1\rangle \rightarrow \langle n, d_2 \rangle$ from \emph{PathWorkList} \\
            \textbf{let} $f = JumpFn(\langle s_p, d_1\rangle \rightarrow \langle n, d_2 \rangle)$ \\
            \Switch{
                $(n)$
            }{
                \Case{
                    $n$ is a call node in $p$, calling a procedure $q$
                }{
                    \For{
                        $d_3$ s.t. $\langle n, d_2\rangle \rightarrow \langle s_q, d_3 \rangle \in E^\#$
                    }{
                        \FPropagate{$\langle s_q, d_3\rangle \rightarrow \langle s_q, d_3 \rangle, id$}
                    }
                    \textbf{let} $r$ be the return-site node that corresponds to $n$ \\
                    \For{
                        $d_3$ s.t. $e = \langle n, d_2\rangle \rightarrow \langle r, d_3 \rangle \in E^\#$
                    }{
                        \FPropagate{$\langle s_p, d_1\rangle \rightarrow \langle r, d_3 \rangle, EdgeFn(e) \circ f$}
                    }
                    \For{
                        $d_3$ s.t. $f_3 = SummaryFn(\langle n, d_2\rangle \rightarrow \langle r, d_3 \rangle) \neq \lambda l.\top$
                    }{
                        \FPropagate{$\langle s_p, d_1\rangle \rightarrow \langle r, d_3 \rangle, f_3 \circ f$}
                    }
                }
                \Case{
                    $n$ is the exit node of $p$
                }{
                    \For{
                        call node $c$ that calls $p$ with corresponding return-site node $r$
                    }{
                        \For{
                            $d_4$, $d_5$ s.t. $\langle c, d_4\rangle \rightarrow \langle s_p, d_1\rangle \in E^\#$ and $\langle e_p, d_2\rangle \rightarrow \langle r, d_5\rangle \in E^\#$
                        }{
                            \textbf{let} $f_4 = EdgeFn(\langle c, d_4\rangle \rightarrow \langle s_p, d_1\rangle)$ and \\
                            $f_5 = EdgeFn(\langle e_p, d_2\rangle \rightarrow \langle r, d_5\rangle)$ and \\
                            $f' = (f_5 \circ f \circ f_4) \sqcap SummaryFn(\langle c, d_4\rangle \rightarrow \langle r, d_5\rangle)$
                        }
                        \uIf{
                            $f' \neq SummaryFn(\langle c,d_4 \rangle \rightarrow \langle r, d_5 \rangle)$
                        }{
                            $SummaryFn(\langle c,d_4 \rangle \rightarrow \langle r, d_5 \rangle) \coloneqq f'$ \\
                            \textbf{let} $s_q$ be the start node of $c$'s procedure \\
                            \For{
                                $d_3$ s.t. $f_3 = JumpFn(\langle s_q, d_3 \rangle \rightarrow \langle c, d_4 \rangle) \neq \lambda l.\top$
                            }{
                                \FPropagate{$\langle s_q, d_3 \rangle \rightarrow \langle r, d_5 \rangle, f' \circ f_3$}
                            }
                        }
                    }
                }
                \Case{
                    $n$ is an intraprocedural node in $p$
                }
                {
                    \For{
                        $ \langle m, d_3 \rangle  s.t. \langle n, d_2 \rangle \rightarrow \langle m, d_3 \rangle \in E^\# $
                    }{
                        \FPropagate{$\langle s_p, d_1 \rangle \rightarrow \langle m, d_3 \rangle$,\\
                        $\qquad\qquad\qquad EdgeFn(\langle n, d_2 \rangle \rightarrow \langle m, d_3 \rangle) \circ f$}

                    }
                }
            }
        }
    }
    \;
    \Pn{\FPropagate{e, f}}{
        \textbf{let} $f' = f \sqcap JumpFn(e)$ \\
        \If{
            $f' \neq JumpFn(e)$
        }{
            $JumpFn(e) \coloneqq f'$ \\
            Insert $e$ into \emph{PathWorkList}
        }
    }
    \captionsetup{skip=5pt,belowskip=-10pt}
    \caption{The original IDE algorithm for Phase \rom{1} (reproduced from \cite{ide}).}
    \label{ide-p1}
\end{algorithm}

\textbf{Initialization.} In lines 2--5, jump and summary functions are initialized.
Jump functions, denoted by $JumpFn$, correspond to the \emph{same-level realizable paths} (SLRPs) from the start node $s_p$ of a procedure $p$ to a node $m$ in $p$. 
Summary functions, denoted by $SummaryFn$, summarize the effect of a procedure call through same-level realizable paths from the call-site $c$ to return-site $r$.
In line 3, $JumpFn(\langle s_p, d'\rangle \rightarrow \langle m, d\rangle) = \lambda l.\top$ states that the jump function from the node $\langle s_p, d'\rangle$ to each $\langle m, d\rangle$ is 
initialized to $\lambda l.\top$.
In line 5, $SummaryFn(\langle c, d'\rangle \rightarrow \langle r, d\rangle) = \lambda l.\top$ states that the summary function from each call-site node $\langle c, d'\rangle$ to its corresponding return-site $\langle r, d\rangle$ is initialized to $\lambda l.\top$.
Line 6 initializes the \emph{PathWorkList} to $\{\langle s_{main}, \Lambda\rangle\ \rightarrow \langle s_{main}, \Lambda\rangle \}$ representing a self-loop edge on the start node of the \emph{main} procedure whose jump function is the identity function, \emph{id}.
The jump function from the start node $s_p$ until the current statement $n$ is denoted with $f$.

\textbf{Call nodes.} Lines 12-19 handle the case where $n$ is a call-site node in $p$, calling a procedure $q$.
In line 14, the self-loop edge on the start node of the callee procedure $q$ is initialized with \emph{id}.
In line 17, the edge from $s_p$ the corresponding return-site $r$ is computed by composing the $f$, the jump function until $n$ and the edge function from $n$ to $r$.
In line 19, the edge from $s_p$ the corresponding return-site $r$ is computed by composing $f$ and $f_3$, the corresponding summary function when it is not mapping to $\top$.

\textbf{Exit nodes.} Lines 20-30 handle the case where $n$ is the exit node of $p$.
Edges from each call-site node $c$ to the start node $s_p$ (shown with $f_4$) and from the exit node, $e_p$ to each caller's return-site $r$ (shown with $f_5$) must be computed. In line 25, a new summary function $f'$ is computed by composing $f_5$, $f$, and $f_4$ and merging the existing summary function for the same $c$ and $r$. When it is a new summary, a new jump function is computed from the caller procedure's start node $s_q$ to the node return-site node $r$ by composing the $f'$ with the existing jump function $f_3$ from $s_q$ to call-site node $c$.

\textbf{Normal nodes.} Lines 31-33 handle the case where $n$ is a non-call or intraprocedural node.
Edges from the start node $s_p$ to each node $m$, which is the statement that appears directly after $n$ in procedure $p$, are computed by composing the edges from $s_p$ to $n$ (shown with $f$) and the edges from $n$ to $m$.

\LinesNumbered
\DontPrintSemicolon
\begin{algorithm}[t]
    \setstretch{1.35}
    \SetInd{0em}{0.5em}
    \scriptsize
    \SetKwFunction{FMain}{ForwardComputeSparseJumpFunctionsSLRPs}
    \SetKwFunction{FPropagate}{Propagate}
    \SetKwFunction{FNextUse}{NextUse}
    \SetKwData{Variable}{variable}
    \SetKwData{Value}{value}
    \SetKwProg{Fn}{Function}{:}{}
    \SetKwProg{Pn}{Function}{:}{}
    \Pn{\FMain{}}{
        \ldots \\
        \setcounter{AlgoLine}{7}
        \While{
            PathWorkList $ \neq \emptyset $
        }{
            Select and remove an item $\langle s_p, d_1\rangle \rightarrow \langle n, d_2 \rangle$ from \emph{PathWorkList} \\
            \textbf{let} $f = JumpFn(\langle s_p, d_1\rangle \rightarrow \langle n, d_2 \rangle)$ \\
            \Switch{
                $(n)$
            }{
                \Case{
                    $n$ is a call node in $p$, calling a procedure $q$
                }{
                    \ldots \\
                    \setcounter{AlgoLine}{14}
                    \textbf{let} $r$ be the return-site node that corresponds to $n$ \\
                    \For{
                        $d_3$ s.t. $e = \langle n, d_2\rangle \rightarrow \langle r, d_3 \rangle \in E^\#$
                    }{
                        \HiLi\textbf{let} $r'=$ \FNextUse{$p, d_3, r$} \\
                        \HiLi\FPropagate{$\langle s_p, d_1\rangle \rightarrow \langle r', d_3 \rangle$,\\
                        \HiLi$\qquad\qquad\qquad EdgeFn(\langle n, d_2 \rangle \rightarrow \langle r, d_3 \rangle) \circ f$}
                    }
                }

                \ldots \\
                \setcounter{AlgoLine}{30}
                \Case{
                    $n$ is an intraprocedural node in $p$
                }{
                    \For{
                        $ \langle m, d_3 \rangle  s.t. \langle n, d_2 \rangle \rightarrow \langle m, d_3 \rangle \in E^\# $
                    }{
                        \HiLi\textbf{let} $m'=$ \FNextUse{$p, d_3, n$} \\
                        \HiLi\FPropagate{$\langle s_p, d_1 \rangle \rightarrow \langle m', d_3 \rangle$,\\
                        \HiLi$\qquad\qquad\qquad EdgeFn(\langle n, d_2 \rangle \rightarrow \langle m, d_3 \rangle) \circ f$}
                    }
                }
            }
        }
    }
    \;
    \setcounter{AlgoLine}{40}
    \Pn{\FNextUse{p, d, n}}{
        \textbf{let} $G_{p,d}$ be the sparse CFG of $d$ in procedure $p$ \\
        \textbf{let} $C$ be the sparse CFG cache with $(p,d)$ typed keys and $G_{p,d}$ as values \\
        \If{
            $G_{p,d} \notin$ $C$
        }{
            construct $G_{p,d}$ and add to $C$
            
        }
        \KwRet{the next statement after $n$ from $G_{p,d}$}
    }
    \captionsetup{skip=5pt,belowskip=-14pt}  
    \caption{Modifications for Sparse IDE algorithm for Phase \rom{1} (mirrors the design from \cite{sparsedroid}).
    }
    \label{side-p1}
\end{algorithm}

\subsection{The Sparse IDE Algorithm}
In the original IDE algorithm, each symbol $d \in D \cup \{\Lambda\}$ at a statement $n$ is propagated to its direct successor statement $m$. As also pointed out in previous work~\cite{sparsedroid}, this behavior is desired when $n$ is a call and exit node. For these nodes, the reachability of each $d$ in different contexts is left to the data-flow function definition. \emph{call-flow functions} propagate each $d$ into the context of the callee procedure. \emph{return-flow functions} propagate each $d$ back to the context of the caller procedure. \emph{call-to-return-flow functions} propagate each $d$ from before a procedure is called to after the procedure is called. However, when $n$ is a non-call node, each $d$ can safely be propagated to $d$'s next use statement.

Figure \ref{side-p1} shows the modifications for the Sparse IDE algorithm for Phase \rom{1}.
We replace line 17 from the original IDE algorithm with lines 17-19 in the Sparse IDE algorithm. 
Instead of propagating $d_3$ to the direct return site node $r$, we obtain $r'$ which is the next use statement of $d_3$ in its \emph{symbol-specific} sparse control flow graph. 
Similarly, we replace line 33 with lines 33-35, to propagate $d_3$ to its next use statement $m'$ its sparse control flow graph.
Our sparsification approach mirrors that of sparse IFDS algorithm \cite{sparsedroid}, however, since we generalize it to IDE, we also account for edge function composition.

\subsection{Sparse IFDS Revisited}\label{cond}
As shown in Figure~\ref{fig:sparseifds}, a statement can behave as \emph{identity function}, meaning it does not affect any data-flow fact, $d \in D$. 
However, as shown by He et al. \cite{sparsedroid}, many statements only affect a few data-flow facts, 
often even just a single fact. Their flow functions can be considered  \emph{fact-specific identity functions} for the facts that they do not affect. Sparse IFDS defines fact-specific identity functions as follows \cite{sparsedroid}: 

Given a symbol, $d \in D$ and a flow function, $f \in 2^D \rightarrow 2^D$, $f$ is a \emph{d-specific identity function} if the following conditions hold:

    \begin{align}
        \forall X \in 2^D &: d \in X \Rightarrow d \in f(X) \tag{1.1}\label{c11} \\
        \forall X \in 2^{D \setminus \{d\}} &: f(X) \setminus \{d\} = f(X \cup \{d\}) \setminus \{d\} \tag{1.2}\label{c12}
    \end{align} \label{eq:1}

Condition \ref{c11} states that $d$ is not affected by other facts when applying $f$, and \ref{c12} states that $d$ does not affect the other facts when applying $f$. However, these conditions only apply to symbols from $D$ and ignore mappings from $D$ to the value domain $L$, and, if applied to IDE problems, one would wrongly treat such flow functions that are annotated with non-identity edge functions as $d$-specific identity functions as well.

\begin{figure}[tb!]
    \centering
    \includegraphics[width=\linewidth,keepaspectratio]{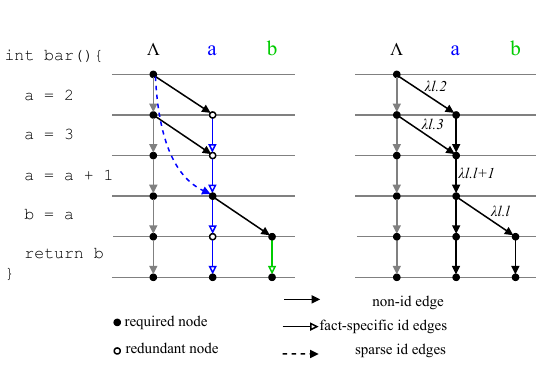}
    \subfloat[\label{fig:theirs} Sparse IFDS]{\hspace{.325\textwidth}}
    \subfloat[\label{fig:ours} Sparse IDE]{\hspace{.125\textwidth}}
    \captionsetup{skip=5pt,belowskip=-10pt}
    \caption{Comparison of the Sparsification Approaches of Sparse IFDS and Sparse IDE}
    \label{fig:theirs-ours}
\end{figure}

Figure \ref{fig:theirs-ours} shows two important cases where sparse IFDS would sparsify incorrectly. First, reassignments: \texttt{a~=~3} reassigns \texttt{a}, but sparse IFDS recognizes that $a$ already exists (is ``tainted''), and therefore it treats this statement as $a$-specific identity. Second, value updates: \texttt{a~=~a~+~1} updates $a$'s value, but sparse IFDS has no notion of values, therefore, from its perspective, this statement is ``identity'' as well. Sparse IDE, on the other hand, is aware of the effects on the value domain and retains both statements.

\subsection{Fact-Specific Identity Transformers}
To generalize fact-specific sparsification to the IDE framework, we define symbol-specific identity transformers that take into account the environments that map the symbols from domain $D$ to the values from domain $L$. 
Given a symbol $d \in D$ and a value $l \in L$, $env~=~[d \mapsto l]$ is an environment $env$ mapping from $d$ to $l$, i.e., $env(d)=l$. Then $env$ is an element of the set of environments $Env(D, L)$. An environment transformer, $t \in Env(D,L)~\rightarrow~Env(D,L)$ is a \emph{$d$-specific identity transformer}, denoted by $t \equiv t^d$, if the following holds:

First, the transformer $t$ keeps all $d$-specific mappings intact:
\begin{align*}
    \text{given}~ d \in D: \forall env \in Env(D,L):\\
     env(d) = t(env(d)) \tag{2.1}\label{c21}
\end{align*}
Second, for all other mappings, $t$ produces identical results no matter whether or not $d$-specific mappings are present:

\begin{align*}
    \text{given}~ d \in D: \forall env \in Env(D,L).~\forall d' \in D \setminus \{d\}.~\forall l \in L:\\
     t(env(d')) = t(env[d \mapsto l](d')) \tag{2.2}\label{c22}
\end{align*}

We test the edge functions from Figure \ref{fig_ef} on these conditions. 
$e_{id}$ is an $a$-specific identity transformer ($e_{id} \equiv e_{id}^a$), because applying $\lambda env.env$ does not change a's previous mapping. 
$e_{val}$ is not an $a$-specific identity transformer ($e_{val} \not\equiv e_{val}^a$), because applying $\lambda env.env[a \mapsto 3]$ changes a's previous mapping. 
$e_{op}$ is also not an $a$-specific identity transformer ($e_{op} \not\equiv e_{op}^a$) because applying $\lambda env.env[b \mapsto 2 * env(a) + 1]$ changes another value's mapping (for $b$) depending on what $a$ maps to, and because it changes b's value $e_{op}$ is not a $b$-specific identity transformer either ($e_{op} \not\equiv e_{op}^b$).
Note that, importantly, a transformer can only be considered a $d$-identity transformer if the above restrictions hold \emph{irrespective} of any concrete $l\in L$ that might be associated with $b$: (\ref{c22}) quantifies over all $l \in L$. This is necessary because IDE produces procedure summaries that must be sound with respect to all $l$, and thus their creation must not be made dependent on $l$. In other words, IDE can support symbol-specific but not value-specific sparsification!

\subsection{Determining symbol-specific identity}
When propagating fact $d$, we consider only those statements as irrelevant statements for $d$ that fulfil conditions (\ref{c21}) and (\ref{c22}). But since these conditions are value-agnostic---they quantify over all $l\in L$, this allows one to determine \emph{ahead of time} the statements whose environment transformers adhere to both conditions, structurally.
First, by Condition \ref{c21}, a statement's corresponding environment transformer $t$ is \emph{not} a $d$-specific identify transformer if $t$ affects $d$'s value mapping in any way, i.e., $t = \lambda env.env[\bm{d} \mapsto \text{\textunderscore}]$.
Second, by Condition \ref{c22}, $t$ is \emph{not} a $d$-specific identity transformer either, if $t$ uses $d$'s value mapping $env(d)$ to compute another fact's value, i.e. $t = \lambda env.env[\text{\textunderscore} \mapsto \ldots env(\bm{d})\ldots]$. 

Naturally, sparsification effectiveness is closely tied to the analysis-specific environment-transformer definitions. The environment transformer for the statement \texttt{a = a + 1} is $t \equiv t^a$ for taint analysis, where $t=\lambda env.env$. For constant propagation analysis, however, $t \not\equiv t^a$, where $t=\lambda env.env[env(a)+1]$.

Sparse IDE strictly generalizes Sparse IFDS as presented in SparseDroid. One can easily define sparse IFDS as an instantiation of sparse IDE by restricting the value domain $L$ to $\{\bot,\top\}$, where symbols that map to $\bot$ are considered reachable. In this setting, our definitions (\ref{c21}) and (\ref{c22}) become equivalent to (\ref{c11}) and (\ref{c12}).

\section{Application to Linear Constant Propagation} \label{sec:impl}
As Sagiv, Reps and Horwitz explain in their seminal work \cite{ide}, 
constant propagation analysis is the perfect problem setting where IDE outperforms IFDS \cite{ifds}.
This is not only because the problem's lattice is larger than the binary domain, 
but also it is infinitely broad where IFDS cannot terminate.
We are, therefore, motivated to apply the Sparse IDE framework to linear constant propagation analysis. 
\myh{}, and thus \mysh{}, are generic tools and they are independent of the target language and their intermediate representations (IRs). In this work, we use \mysoot{} \cite{soot} static program analysis framework for Java and its intermediate representation \myjimple{}. Therefore, in the following, we explain our implementation based on the \myjimple{} IR.

\begin{table*}[!hbt]
    \renewcommand{\arraystretch}{1.8}
    \scriptsize
    \caption{Statements for Linear Constant Propagation Analysis with Corresponding IRs and Flow/Edge Functions.}
    \centering
    \begin{tabular}{llll}
    \toprule
    Statement & IR & Flow Function & Edge Function \\
    \midrule
    \rowcolor{gray!15} constant  &  $a \leftarrow Const$ & $ \;\;\; \lambda S.\{S \cup \{a\}\}$ & $\lambda env.env[a \mapsto Const]$ \\
    binop & $a \leftarrow b \odot Const$ & 
    \raggedright
    \begin{minipage}{0.146\linewidth}
     \begin{equation*}
        \lambda S. 
        \begin{cases}
            \begin{aligned}
            S \: \cup \: &\{a\} \quad \text{if } b \in S \\
            S \: \setminus \: &\{a\}
            \end{aligned}
        \end{cases}
    \end{equation*}
    \end{minipage}
    
    &  $\lambda env.env[a \mapsto env(b) \, \hat{\odot} \, Const]$ \\
    \cellcolor{gray!15}
    local     & \cellcolor{gray!15} $a \leftarrow b$       &
    \raggedright
    \cellcolor{gray!15}
    \begin{minipage}{0.14\linewidth}
    \begin{equation*}
        \lambda S. 
        \begin{cases}
            \begin{aligned}
            S \: \cup \: &\{a\} \quad \text{if } b \in S \\
            S \: \setminus \: &\{a\}
            \end{aligned}
        \end{cases}
    \end{equation*}
\end{minipage}

    & \cellcolor{gray!15} $\lambda env.env[a \mapsto env(b)]$ \\
    field load      &  $a \leftarrow b.f$   & 
    \raggedright
    \begin{minipage}{0.157\linewidth}
    \begin{equation*}
        \lambda S. 
        \begin{cases}
            \begin{aligned}
            S \: \cup \: &\{a\} \quad \text{if } b.f \in S \\
            S \: \setminus \: &\{a\}
            \end{aligned}
        \end{cases}
    \end{equation*}
\end{minipage}
    
    & $\lambda env.env[a \mapsto env(b.f)]$ \\

    \cellcolor{gray!15} field store     &  \cellcolor{gray!15} $a.f \leftarrow b$   & 
    \raggedright
    \cellcolor{gray!15}
    \begin{minipage}{0.243\linewidth}
    \begin{equation*}
        \lambda S. 
        \begin{cases}
            \begin{aligned}
            S \: \cup \: &\{p.f \mid p \in aliases(a)\} \quad \text{if } b \in S \\
            S \: \setminus \: &\{p.f \mid p \in aliases(a)\}
            \end{aligned}
        \end{cases}
    \end{equation*}
\end{minipage}
    
    & \cellcolor{gray!15} $\lambda env.env[p.f \mapsto env(b)]$  \\

    static field load &  $a \leftarrow T.f$   & 
    \raggedright
    \begin{minipage}{0.157\linewidth}
    \begin{equation*}
        \lambda S. 
        \begin{cases}
            \begin{aligned}
            S \: \cup \: &\{a\} \quad \text{if } T.f \in S \\
            S \: \setminus \: &\{a\}
            \end{aligned}
        \end{cases}
    \end{equation*}
\end{minipage}
    & $\lambda env.env[a \mapsto env(T.f)]$   \\
    \cellcolor{gray!15}
    static field store     & \cellcolor{gray!15}  $T.f \leftarrow b$  & 
    \raggedright
    \cellcolor{gray!15}
    \begin{minipage}{0.243\linewidth}
    \begin{equation*}
        \lambda S. 
        \begin{cases}
            \begin{aligned}
            S \: \cup \: &\{p.f \mid p \in aliases(T)\} \quad\text{if } b \in S \\
            S \: \setminus \: &\{p.f \mid p \in aliases(T)\}
            \end{aligned}
        \end{cases}
    \end{equation*}
\end{minipage}
    & \cellcolor{gray!15} $\lambda env.env[p.f \mapsto env(b)]$  \\

    array load      &  $a \leftarrow A[i]$  & 
    \raggedright
    \begin{minipage}{0.16\linewidth}
    \begin{equation*}
        \lambda S. 
        \begin{cases}
            \begin{aligned}
            S \: \cup \: &\{a\} \quad \text{if } A[i] \in S \\
            S \: \setminus \: &\{a\}
            \end{aligned} 
        \end{cases}
    \end{equation*}
\end{minipage}
    & $\lambda env.env[a \mapsto env(A[i])]$   \\
    \cellcolor{gray!15}
    array store     & 
    \cellcolor{gray!15} $A[i] \leftarrow b$  & 
    \cellcolor{gray!15}
    \raggedright
    \begin{minipage}{0.243\linewidth}
         \begin{equation*}
        \lambda S. 
        \begin{cases}
            \begin{aligned}
            S\: \cup \:&\{p[i] \mid p \in aliases(A)\} \quad\text{if } b \in S \\
            S\: \setminus \:&\{p[i] \mid p \in aliases(A)\} 
            \end{aligned}
        \end{cases}
    \end{equation*}
\end{minipage}
    & \cellcolor{gray!15} $\lambda env.env[p[i] \mapsto env(b)]$  \\

    call    &  $r \leftarrow b.m(a_i) $      & 
    \raggedright
    \begin{minipage}{0.235\linewidth}
        \raggedright
    \begin{equation*}
        \raggedright
        \lambda S. 
        \begin{cases}
            \begin{aligned}
            S\: \cup \:&\{p_i\} \quad \text{if } a_i \in S \land a_i \mapsto p_i \text{ in } m \\
            S\: \setminus \:&\{p_i\}
            \end{aligned}
        \end{cases}
    \end{equation*}
\end{minipage}
    &  $\lambda env.env[p_i \mapsto env(a_i)]$ \\
    \cellcolor{gray!15}
    return    & \cellcolor{gray!15}  $r \leftarrow b.m(a_i) $      & 
    \cellcolor{gray!15}
    \begin{minipage}{0.212\linewidth}
    \begin{equation*}
        \raggedright
        \lambda S. 
        \begin{cases}
            \begin{aligned}
            S\: \cup \:&\{r\} \quad \text{if } r' \in S \land m \text{  returns } r' \\
            S\: \setminus \:&\{r\}
            \end{aligned}
        \end{cases}
    \end{equation*}
\end{minipage}
    & 
    \cellcolor{gray!15}
    $\lambda env.env[r \mapsto env(r')]$ \\

    call-to-return    &  $r \leftarrow b.m(a_i) $      & 
    \raggedright
    \begin{minipage}{0.235\linewidth}
    \begin{equation*}
        \lambda S. 
        \begin{cases}
            \begin{aligned}
            S\: \setminus \:&\{a_i\} \quad \text{if } a_i \in S \land a_i \mapsto p_i \text{ in } m \\
            S \,\quad& 
            \end{aligned}
        \end{cases}
    \end{equation*}
    \end{minipage}

    &  $\lambda env.env$ \\

    \bottomrule
\end{tabular}
    \label{tab:cpa}
    \end{table*}

\subsection{Analysis Definition}\label{sec:analysisdef}
Linear constant propagation analysis handles the linear expressions that generate a new data-flow fact by using just a single other fact, e.g. \texttt{a = b} or \texttt{a~=~2*b~+~1}.
Full constant propagation analysis involves statements such as \texttt{a~=~b~+~c}.
Such a statement's flow function is not distributive; it cannot be precisely computed within the IDE framework. Our linear constant propagation analysis implementation handles the assignment statements shown in Table~\ref{tab:cpa}.

\textbf{IR}. The IR always ensures binary operation (\emph{binop}) representation by reducing more complex operations to binary operations.
For instance, \texttt{a~=~2*b~+~1} would be reduced to \texttt{s1~=~2~*~b} and \texttt{a~=~s1~+~1}. The IR also reduces longer access paths to multiple assignments with a single access path (\emph{n=1}). For instance, a statement such as \texttt{a = b.f1.f2} would be reduced to \texttt{s1~=~b.f1}, \texttt{s2~=~s1.f2}, and \texttt{a~=~s2}. 
The same reduction applies to procedure invocations as well.

\textbf{Flow functions.}
We \emph{generate} a symbol when it is assigned with a \emph{constant}. As discussed, we handle the binary operations in the linear form. We distinguish between the assignments that require alias handling and the ones that do not. The assignments such as \emph{local, field load, static field load,} and \emph{array load}, overwrite the local variable, $a$, on their left-hand side and therefore do not need to know $a$'s aliases. The assignments such as \emph{field store, static field store,} and \emph{array store}, on the other hand, require handling the aliases of the base variables or the array references. To handle aliasing we use the \myb{} \cite{boomerang} demand-driven pointer analysis framework. When necessary, we query the aliases of the base variables and add them to the set of propagated symbols. Note that in Table \ref{tab:cpa}, the alias sets contain the query variable as well. The IDE framework requires three types of flow functions to model the effects of invoke statements. The \emph{call} flow function propagates the symbol for the actual parameter to the context of the callee procedure, by mapping it to the procedure's corresponding formal parameter. The \emph{return} flow function propagates the symbol for the returned variable to the context of the caller procedure, by mapping it to the symbol on the left-hand side of the invoke expression. The \emph{call-to-return} flow function propagates the symbols that are not passed to the context of the callee procedure, to the next statement after the invoke statement. 

\textbf{Edge functions.}
For most statements, the edge functions map the target symbol to the value of the source symbol, 
acting as \emph{identity transformers}. The \emph{constant} and \emph{binop} statements are the only exceptions. The constant statement maps the target symbol, $a$ to the given constant value, $Const$. The binop statement maps the target symbol, $a$ to a new value. The value is computed by simulating the operation $\odot$ using the source symbol's value, $env(b)$ and the constant operand, $Const$.
Edge functions must be composed and reduced to a simple value mapping when computing the actual values. Given $f_1, f_2 \in Env(D,~L)$ and $f_1$ appears before $f_2$ as an edge in the exploded supergraph, we compose the edge functions as follows:

\small
\begin{equation*}
    f_2 \circ f_1 :=  
    \begin{cases}
        f_2 \qquad  & \text{if}~ f_1 = \lambda env.env \\
        f_1 \qquad  &\text{if}~f_2 = \lambda env.env \\
        f_2 \qquad &\text{if}~f_2 = \lambda env.env[a \mapsto Const] \\
        f_2(f_1) \qquad &\text{if}~f_2 = \lambda env.env[a \mapsto env(b) \, \hat{\odot} \, Const] \\
    \end{cases}
\end{equation*}
\normalsize

If an edge function is the identity transformer, we always apply the other function by the first two conditions. We always apply the subsequent edge function if it is a constant assignment, by the third condition. If the subsequent edge is a binop, we compute its value immediately in place by applying the preceding edge first, as suggested in previous work~\cite{heros}.

\textbf{Lattice.} We perform the linear constant propagation on integers. Therefore the lattice is $\mathbb{Z}^\top_\bot$.
Given $l_1, l_2 \in \mathbb{Z}^\top_\bot$, we define the meet operator as follows:

\small
\begin{equation*}
    l_1 \sqcap l_2 =  
    \begin{cases}
        l_1 \qquad  &\text{if}~l_2=\top \\
        l_2 \qquad &\text{if}~l_1=\top \\
        \bot \qquad &\text{if}~l_1=\bot \; \lor \; l_2=\bot  \\
        \top \qquad &\text{if}~l_1=\top \; \land \; l_2=\top
    \end{cases}
\end{equation*}
\normalsize

If a value is $\top$, the meet operator yields the other value by the first two conditions. If either of the values is $\bot$, the meet yields $\bot$, and if both of the values are $\top$ it yields $\top$ by the third and fourth conditions respectively. 

\subsection{Sparsification for Constant Propagation}
Our sparsification approach has much in common with the one proposed by He et al.~\cite{sparsedroid}, though modifications were necessary. We build the sparse control flow graphs (CFGs) by ignoring symbol-specific identity functions. Given a procedure, $p$, $G_p$ is its original \emph{dense} CFG. We build sparse CFGs specific to each symbol, $d$ in $p$, denoted as $G_{p,d}$ and propagate $d$ across its own sparse CFG. As shown with the IR in Table \ref{tab:cpa}, $d$ can be a local, an instance field or static field, or an array access. $G_{p,d}$ is constructed by determining whether each statement's corresponding flow function in  $G_p$ is a $d$-specific identity function.

As a major modification, and most importantly, we account for a statement's effect on the value domain. In addition to determining whether each statement's corresponding flow function is a \emph{d-specific identity function}, we determine whether its edge function is a \emph{d-specific identity transformer} with the assumptions explained in Section \ref{cond}. 
Further, we propagate the \emph{tautological} fact, $\Lambda$, (sparsely) to the statements that can generate new data-flow facts, e.g. $a~\gets~Const$. Otherwise, it is impossible to generate new facts at arbitrary program points.
Finally, we soundly retain all branching statements to keep the original CFGs' control flow as it is.

\setlength{\belowcaptionskip}{-10pt}
\begin{figure*}[t]
    \centering
    \includegraphics[width=\textwidth]{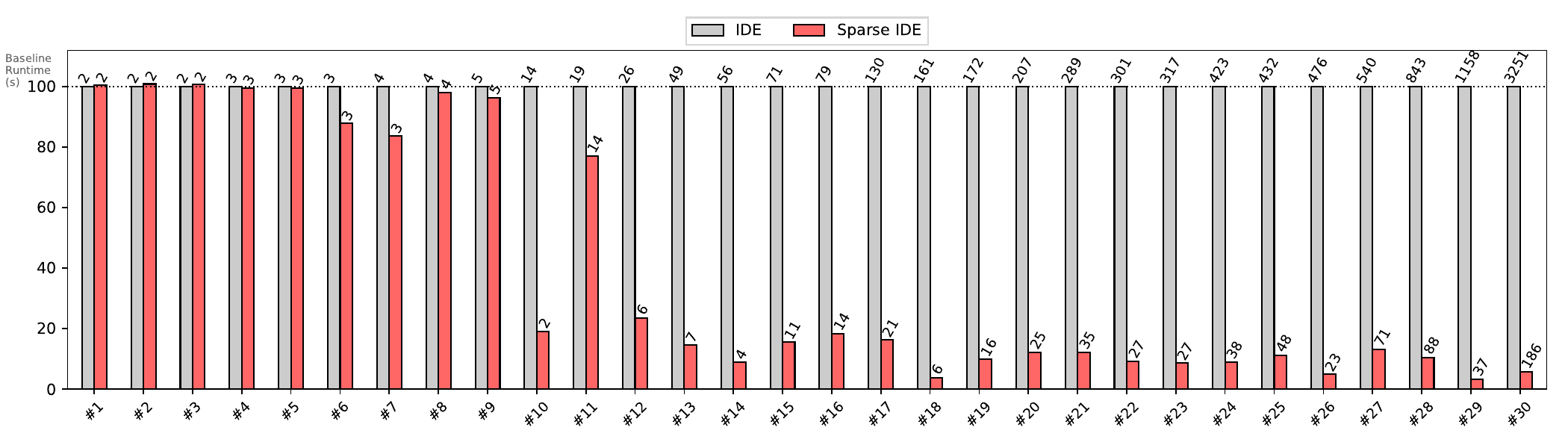}
    \caption{Relative runtime of Sparse IDE compared to the baseline original IDE in \%, annotated with exact runtimes in seconds, sorted by original IDE's runtime}
    \label{fig:runtime}
\end{figure*}

\begin{figure*}[t]
    \centering
    \includegraphics[width=\textwidth]{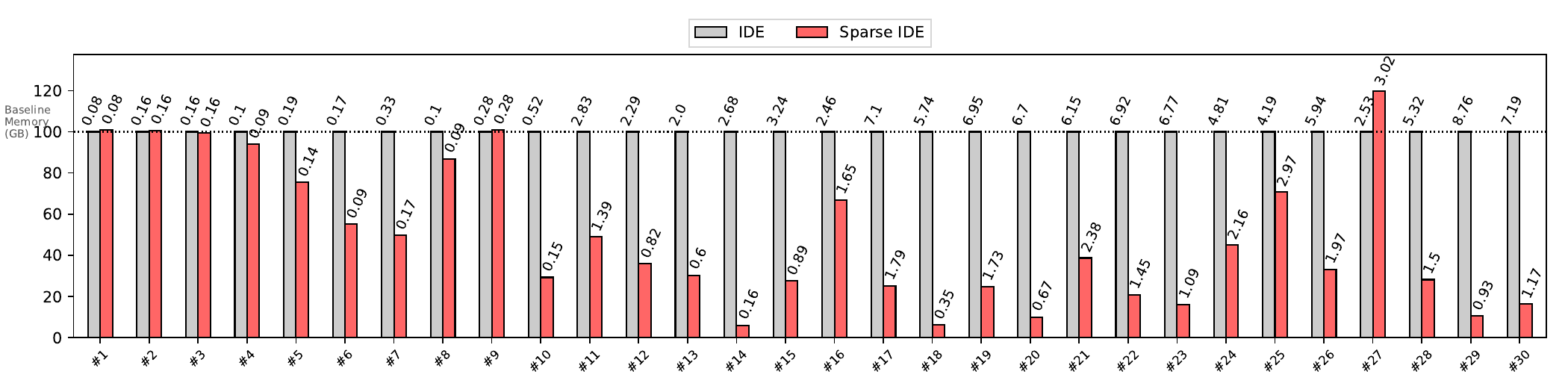}
    \caption{Memory consumption of Sparse IDE compared to the baseline original IDE in \%, annotated with exact memory consumptions in GB, using the same sorting as Figure \ref{fig:runtime}}
    \label{fig:memory}
\end{figure*}

\section{Evaluation} \label{sec:eval}
We next explain the research questions that guide our evaluation and its experimental setup, and then we discuss the evaluation results. Sparse data-flow analyses promise extensive performance improvements,
while still maintaining the precision of their non-sparse counterparts. Therefore, first, we compare the sparse analysis results against the non-sparse analysis results. Second, we measure whether the sparse analysis produces the promised performance benefits. Third, we investigate the factors contributing to the performance impact. Therefore, we focus on the following research questions:

\begin{itemize}
    \item \rqone{}
    \item \rqtwo{}
    \item \rqthree{}
\end{itemize}

\subsection{Experimental Setup}
We implement the proposed approach in \mysh{}, by extending the open source \myh{} IDE solver's latest version, at the time of writing (\texttt{e7e4a85}) \cite{HerosGithub}. Using \mysh{} and the \mysoot{} static analysis framework \cite{soot}, we implement a linear constant propagation analysis.
To handle aliasing, we integrate our client analysis with the \myb{} \cite{boomerang} demand-driven pointer analysis, using its latest version (\texttt{1179227}) \cite{BoomerangGithub}.
\myh{}, and thus \mysh{}, support multi-threading, yet, because \myb{} is single-threaded,
our client analysis uses a single-thread. Therefore, our evaluation results present single-thread performance.

As benchmark subjects we use:

\begin{itemize}
    \item \textbf{\mybench{}}: A benchmark suite for constant propagation analysis targeting Java, did not previously exist. We, therefore, created \mybench{} as a micro-benchmark suite for integer linear constant propagation analysis. 
    We run both \myh{} and \mysh{} on this benchmark suite and compare the analysis results that they produce. 

    \begin{table}
        \footnotesize
        \centering
        \captionsetup{skip=5pt,belowskip=-10pt}
        \caption{\mybench{} Test Cases}
        \begin{tabular}{p{0.48\linewidth}p{0.38\linewidth}}
        \toprule
        \textbf{Assignment}  & \textbf{Field Sensitivity} \\ \hline 
        Constant \vspace{0.1cm} & LoadConstant \\ 
        ConstantBinop \vspace{0.1cm} & StoreConstant \\ 
        LocalBinop \vspace{0.1cm} & StoreViaAlias \\ 
        LocalMultipleBinop \vspace{0.1cm} & StoreBinop \\ 
        Overwrite \vspace{0.1cm} & FieldToField \\
        Increment \vspace{0.1cm} & StoreBinopViaAlias \\ 
        Operators  & StoreLocalViaAlias \\
        \cmidrule(l{0.52\linewidth}r{0em}){1-2}
        AssignmentChain    &  \textbf{Context Sensitivity} \\ 
        \cmidrule(l{0.52\linewidth}r{0em}){1-2}
        Static & Id \\
        \cmidrule(l{0em}r{0.46\linewidth}){0-1}
        \textbf{Branching} &  Increment \\
        \cmidrule(l{0em}r{0.46\linewidth}){0-1}
        SameValueMergedAndUsed \vspace{0.1cm}& Add \\
        SameValueMergedNotUsed \vspace{0.1cm} & Nested \\ 
        SameValueMergedAndUsedInBinop \vspace{0.1cm}  & AssignFieldInCallee \\ 
        DiffValuesMergedAndUsed  & AssignStaticInCallee \\
        \cmidrule(l{0.52\linewidth}r{0em}){1-2}
        DiffValuesMergedNotUsed & \textbf{Array} \\
        \cmidrule(l{0.52\linewidth}r{0em}){1-2}
        DiffValuesMergedAndUsedInBinop & LoadConstant \\
        \cmidrule(l{0em}r{0.46\linewidth}){0-1}
        \textbf{Loops} & StoreConstant \\
        \cmidrule(l{0em}r{0.46\linewidth}){0-1}
        ForLoopFixedBound \vspace{0.1cm} & ArrayToArray \\
        ForLoopUnkownBound \vspace{0.1cm} & AliasedArrays \\ 
        WhileTrue  & LargeIndex \\
        \cmidrule(l{0.52\linewidth}r{0em}){1-2}
        WhileUnknown & \textbf{Non-Linear} \\
        \cmidrule(l{0.52\linewidth}r{0em}){1-2}
        NestedLoops \vspace{0.1cm} & Binop \\ 
         & HashCode \\
        \bottomrule
        \end{tabular}
        \label{tab:bench}
    \end{table}

    \item \textbf{Real-world Libraries}\label{selection}: We include real-world Java libraries to investigate the performance of our approach under the workload of large-scale and complex programs. 
    As opposed to applications, libraries do not have a specific \emph{entry} method.
    We follow the \emph{closed package assumption} \cite{library_analysis} for analyzing library code, and treat public methods of the libraries as entry methods.
    We consider a method as an entry method if it adheres to the following entry method selection criteria:
    \begin{itemize}
        \item \textbf{c1:} The method is a public instance method that is not abstract, native or a constructor,
        \item \textbf{c2:} The method contains an integer assignment statement.
    \end{itemize}
    We selected the most downloaded (>5000) Java libraries from the maven repository \cite{MavenRepo}.
    We discarded the libraries that do not contain any entry methods according to the selection criteria, 
    and the ones that caused an error in the underlying static analysis tool, \mysoot{} \cite{soot}.
    In the end, we retained 30 libraries.
    \item \textbf{Replication Package}:
    We set up a replication package, available at \url{https://zenodo.org/records/10461449}
\end{itemize}

We have performed the evaluations on an Intel i7 Quad-Core at 2,3 GHz with 32GB memory.
We configured the JVM with 25GB maximum heap size (\texttt{-Xmx25g}) and 1GB stack size (\texttt{-Xss1g}). 

\subsection{\rqone{}}
\mybench{} consists of 40 target programs with various program properties and sensitivity-testing edge cases, as listed in Table \ref{tab:bench}. \emph{Assignment} cases test possible flow and edge functions, as well as flow sensitivity.
\emph{Branching} and \emph{Loops} cases test the meet operation. \emph{Field sensitivity} cases test field sensitivity and aliasing scenarios.  \emph{Context sensitivity} cases test various calling contexts. \emph{Array} cases test array handling and \emph{NonLinear} cases test analysis' behavior under unanticipated non-linear operations. The results validate the correctness of Sparse IDE by showing that \mysh{} produces the same outputs as the non-sparse \myh{}. 

\subsection{\rqtwo{}}

 Figure \ref{fig:runtime} shows the relative analysis runtime spent by Sparse IDE in comparison to the runtime of the baseline original IDE algorithm. 
We sorted the results for each library by the time spent by the original IDE algorithm. Note that we keep the same sorting for the rest of the paper.
This sorting highlights the fact that our Sparse IDE approach pays off better for the cases where the original IDE's runtime is relatively larger. 
Sparse IDE, compared to the original IDE algorithm, performs up to 30.7x faster.
We measure the mean speedup as 7.9x, and the median speedup as 6.7x. The concrete measurements are presented in Table \ref{tab:results}. 
Results show that, in terms of runtime, Sparse IDE outperforms the original IDE in each run, 
except for the libraries \#1-\#3 (jcl-over-slf4j, slf4j-api, lombok), which have the shortest analysis time. 
In each run, Sparse CFG construction overhead is lower than 1\% of the Sparse IDE total analysis runtime, which is substantially smaller than the achieved speedups.

    Figure \ref{fig:memory} shows the relative memory consumption of Sparse IDE in comparison to the memory consumption of the original IDE algorithm. 
    We have measured up to 94\% reduction in memory consumption in the best case,
    and up to a 19\% increase in the worst. 
    The Sparse IDE algorithm, compared to the original IDE, associates data-flow facts with fewer statements, therefore, we anticipated memory improvements.
    On the other hand, because we cache sparse CFGs ($G_{d,p}$) per each symbol and procedure pair ($d,p$), for some input programs memory consumption increases. 
    However, as shown in Figure~\ref{fig:memory}, these cases are limited to a few outliers.
    Moreover, the mean and median impacts on memory consumption are 51\% and 63\% reduction, respectively.

We statistically assess the significance of the Sparse IDE algorithm's impact on runtime and memory improvements. According to Wilcoxon signed-rank test \cite{wilcoxon} at 0.05 significance level, Sparse IDE significantly improves both the runtime ($p=6.1\mathrm{e}{-08}$) and memory consumption ($p=5.7\mathrm{e}{-07}$) of the original IDE algorithm.

\begin{figure}[t]
    \centering
    \includegraphics[width=0.9\linewidth]{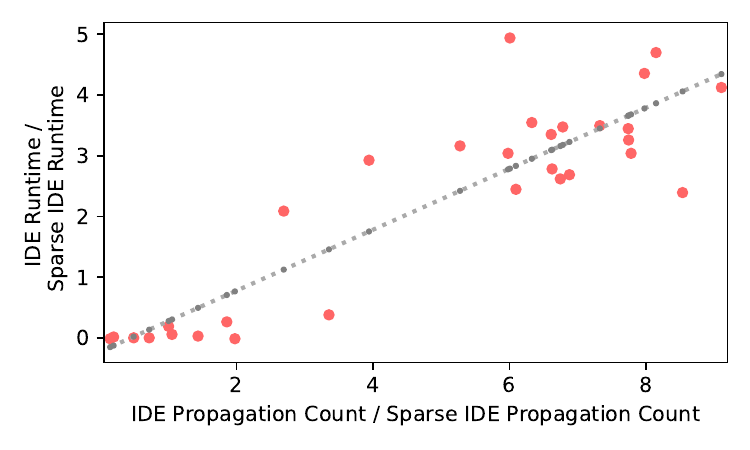}
    \captionsetup{skip=5pt,belowskip=-10pt}
    \caption{Ratio of data-flow fact propagations and corresponding speedup ratios, in log scale}
    \label{fig:prop-to-runtime}
\end{figure}

\begin{figure}[t]
    \centering
    \includegraphics[width=0.9\linewidth]{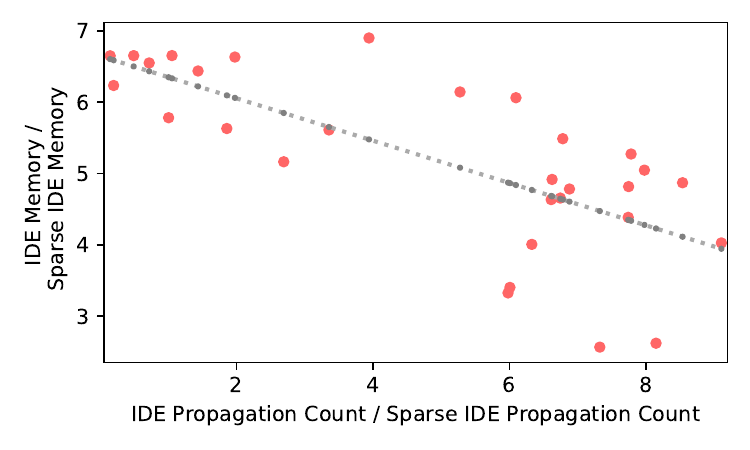}
    \captionsetup{skip=5pt,belowskip=-10pt}
    \caption{Ratio of data-flow fact propagations and corresponding memory consumption ratios, in log scale}
    \label{fig:prop-to-memory}
\end{figure}

\newcolumntype{g}{>{\columncolor{colgray}}r} 
\newcolumntype{k}{>{\columncolor{colgray}}c} 

\newcolumntype{d}{S[table-format=4]} 
\newcolumntype{p}{>{\columncolor{colgray}}S[table-format=4]}

\newcolumntype{q}{S[table-format=4.2]}
\newcolumntype{w}{>{\columncolor{colgray}}S[table-format=4.2]} 
\npthousandsep{,}
\begin{table*}[h]
    \caption{Performance of Sparse IDE compared to the baseline original IDE algorithm}
    \centering
    \scriptsize
    \tabcolsep=0.08cm
    \begin{tabular}{llldppwqqqgggrrr}
        \hline
        \multicolumn{1}{l}{\multirow{2}{*}{\#}} & \multicolumn{1}{c}{\multirow{2}{*}{Library}} & \multicolumn{1}{c}{\multirow{2}{*}{Version}} & \multicolumn{1}{c}{\multirow{1}{*}{\#Entry}} & \multicolumn{3}{k}{Runtime (s)}      & \multicolumn{3}{c}{Memory (GB)} & \multicolumn{3}{k}{\#Propagations}   & \multicolumn{3}{c}{Sparse CFG} \\  
        \multicolumn{1}{c}{} & \multicolumn{1}{c}{} & \multicolumn{1}{c}{} & \multicolumn{1}{c}{Methods} & \multicolumn{1}{g}{IDE} & \multicolumn{1}{g}{SP} & \multicolumn{1}{g}{IDE/SP} &\multicolumn{1}{r}{IDE} & \multicolumn{1}{r}{SP} & \multicolumn{1}{r}{SP/IDE (\%)} & \multicolumn{1}{g}{IDE} & \multicolumn{1}{g}{SP} & \multicolumn{1}{g}{IDE/SP} & \multicolumn{1}{r}{Count} & \multicolumn{1}{r}{Const. (ms)} & \multicolumn{1}{r}{\%Runtime}  \\ \hline
        1 & jcl-over-slf4j & 2.0.7 & 1 & 2 & 2 & 1.00 & 0.08 & 0.08 & 100.78 & 48 & 34 & 1.41 & 2 & 0 & 0.01 \\
        2 & slf4j-api & 2.0.7 & 7 & 2 & 2 & 0.99 & 0.16 & 0.16 & 100.62 & 104 & 94 & 1.11 & 13 & 0 & 0.00 \\
        3 & lombok & 1.18.26 & 5 & 2 & 2 & 0.99 & 0.16 & 0.16 & 99.40 & 894 & 227 & 3.94 & 13 & 0 & 0.02 \\
        4 & commons-logging & 1.2 & 14 & 3 & 3 & 1.00 & 0.10 & 0.09 & 93.87 & \numprint{1509} & 917 & 1.65 & 41 & 0 & 0.00 \\
        5 & junit-jupiter-api & 5.9.2 & 10 & 3 & 3 & 1.01 & 0.19 & 0.14 & 75.39 & 182 & 158 & 1.15 & 20 & 0 & 0.00 \\
        6 & jackson-annotations & 2.14.2 & 79 & 3 & 3 & 1.14 & 0.17 & 0.09 & 55.10 & \numprint{13115} & \numprint{6511} & 2.01 & 190 & 0 & 0.00 \\
        7 & maven-plugin-api & 3.9.1 & 13 & 4 & 3 & 1.20 & 0.33 & 0.17 & 49.61 & \numprint{17353} & \numprint{4780} & 3.63 & 294 & 4 & 0.14 \\
        8 & junit-jupiter-engine & 5.9.2 & 23 & 4 & 4 & 1.02 & 0.10 & 0.09 & 86.81 & \numprint{3204} & \numprint{1181} & 2.71 & 105 & 0 & 0.02 \\
        9 & osgi.core & 8.0.0 & 124 & 5 & 5 & 1.04 & 0.28 & 0.28 & 100.83 & \numprint{58675} & \numprint{28247} & 2.08 & 664 & 7 & 0.15 \\
        10 & jakarta.servlet-api & 6.0.0 & 12 & 14 & 2 & 5.25 & 0.52 & 0.15 & 29.28 & \numprint{126656} & 341 & 371.43 & 33 & 0 & 0.00 \\
        11 & commons-io & 2.11.0 & 178 & 19 & 14 & 1.30 & 2.83 & 1.39 & 48.94 & \numprint{156595} & \numprint{15290} & 10.24 & \numprint{1279} & 116 & 0.78 \\
        12 & commons-codec & 1.15 & 77 & 26 & 6 & 4.25 & 2.29 & 0.82 & 35.90 & \numprint{652560} & \numprint{100866} & 6.47 & 532 & 13 & 0.21 \\
        13 & json & 20230227 & 33 & 49 & 7 & 6.88 & 2.00 & 0.60 & 30.24 & \numprint{1071045} & \numprint{10846} & 98.75 & 407 & 0 & 0.00 \\
        14 & logback-classic & 1.4.7 & 93 & 56 & 4 & 11.28 & 2.68 & 0.16 & 5.92 & \numprint{1286543} & \numprint{8027} & 160.28 & 372 & 12 & 0.24 \\
        15 & logback-core & 1.4.7 & 218 & 71 & 11 & 6.44 & 3.24 & 0.89 & 27.55 & \numprint{1739303} & \numprint{14767} & 117.78 & 925 & 0 & 0.00 \\
        16 & gson & 2.10.1 & 147 & 79 & 14 & 5.45 & 2.46 & 1.65 & 66.93 & \numprint{2009909} & \numprint{29391} & 68.39 & \numprint{1586} & 54 & 0.37 \\
        17 & commons-lang3 & 3.12.0 & 318 & 130 & 21 & 6.14 & 7.10 & 1.79 & 25.22 & \numprint{3418491} & \numprint{31856} & 107.31 & \numprint{1144} & 0 & 0.00 \\
        18 & commons-beanutils & 1.9.4 & 109 & 161 & 6 & 25.97 & 5.74 & 0.35 & 6.15 & \numprint{5855012} & \numprint{20640} & 283.67 & 648 & 2 & 0.04 \\
        19 & mockito-core & 5.3.1 & 235 & 172 & 16 & 10.20 & 6.95 & 1.73 & 24.85 & \numprint{5025407} & \numprint{51374} & 97.82 & \numprint{1663} & 119 & 0.71 \\
        20 & junit-jupiter-params & 5.9.2 & 293 & 207 & 25 & 8.22 & 6.70 & 0.67 & 10.03 & \numprint{6266620} & \numprint{99285} & 63.12 & \numprint{1506} & 109 & 0.43 \\
        21 & assertj-core & 3.24.2 & 334 & 289 & 35 & 8.22 & 6.15 & 2.38 & 38.71 & \numprint{10033236} & \numprint{45563} & 220.21 & \numprint{2418} & 37 & 0.11 \\
        22 & commons-collections4 & 4.4 & 620 & 301 & 27 & 10.90 & 6.92 & 1.45 & 20.91 & \numprint{9140963} & \numprint{42741} & 213.87 & \numprint{1796} & 1 & 0.01 \\
        23 & testng & 7.7.1 & 246 & 317 & 27 & 11.68 & 6.77 & 1.09 & 16.08 & \numprint{9329214} & \numprint{116084} & 80.37 & \numprint{2910} & 15 & 0.06 \\
        24 & joda-time & 2.12.5 & 375 & 423 & 38 & 11.11 & 4.81 & 2.16 & 44.93 & \numprint{15151487} & \numprint{137705} & 110.03 & \numprint{3227} & 69 & 0.18 \\
        25 & guice & 5.1.0 & 336 & 432 & 48 & 8.95 & 4.19 & 2.97 & 70.80 & \numprint{15141525} & \numprint{390634} & 38.76 & \numprint{3918} & 58 & 0.12 \\
        26 & hamcrest-all & 1.3 & 290 & 476 & 23 & 20.48 & 5.94 & 1.97 & 33.10 & \numprint{17953051} & \numprint{71200} & 252.15 & \numprint{1105} & 28 & 0.12 \\
        27 & log4j-core & 2.20.0 & 512 & 540 & 71 & 7.60 & 2.53 & 3.02 & 119.69 & \numprint{18746154} & \numprint{1218580} & 15.38 & \numprint{4666} & 64 & 0.09 \\
        28 & jackson-databind & 2.14.2 & 844 & 843 & 88 & 9.57 & 5.32 & 1.50 & 28.20 & \numprint{35842682} & \numprint{166906} & 214.75 & \numprint{7884} & 5 & 0.01 \\
        29 & okhttp & 4.10.0 & 717 & \numprint{1158} & 37 & 30.69 & 8.76 & 0.93 & 10.58 & \numprint{37431312} & \numprint{581852} & 64.33 & \numprint{5928} & 69 & 0.18 \\
        30 & guava-31.1 & jre & \numprint{1332} & \numprint{3251} & 186 & 17.43 & 7.19 & 1.17 & 16.31 & \numprint{131993565} & \numprint{239589} & 550.92 & \numprint{12200} & 4 & 0.00 \\
        \hline
        \end{tabular}
        \label{tab:results}
        \\
    \end{table*}

\subsection{\rqthree{}}
The essence of the Sparse IDE approach is that, compared to the original IDE algorithm, it propagates data-flow facts to fewer statements.
We investigate to what extent this contributes to improving the scalability of the original IDE algorithm.
Figure \ref{fig:prop-to-runtime}, shows how the ratio of data-flow fact propagations in IDE and Sparse IDE correlate with the ratio of runtime speedups.
We observe that reducing the number of propagations is an effective approach to improving IDE's scalability in terms of runtime.
Similarly, Figure \ref{fig:prop-to-memory} correlates the same with the ratio of memory consumptions in IDE and Sparse IDE.
We observe a comparable trend but not to the same degree.
Given these findings, in the future, one could investigate the potential synergies between our approach and recent approaches that improve the scalability, in particular, in terms of memory \cite{cleandroid, diskdroid}.

\section{Limitations and Threats to Validity} \label{sec:threats}
By definition, Sparse IDE can solve the same data-flow problems as the original IDE framework~\cite{ide}.
It requires data-flow analysis problems to be expressible as distributive environment problems.
Many popular static analyses, such as taint analysis for vulnerability detection \cite{flowdroid} 
or typestate analysis for API misuse detection \cite{emmi2021rapid}, are expressible as distributive environment problems.
Just like other fact-specific sparsification approaches \cite{sparsedroid, sparseBoomerang}, 
Sparse IDE also exploits analysis domain knowledge.
Domain-specific analysis semantics must be correctly encoded 
with flow and edge function definitions within the IDE framework.

Sparse IDE should theoretically lead to a similar performance impact on other data-flow analysis problems where IDE is applicable.
For instance, when performing a typestate analysis, Sparse IDE would safely omit the statements that have no impact on the tracked state.
However, due to space constraints, we were not able to empirically show whether our evaluation results carry over to other analysis problems.

The reported evaluation results might depend on the selected set of Java libraries,
and entry-method selection criteria.
Nevertheless, for real-world library selection, we followed the systematic procedure described in Section \ref{selection}.

To account for variations in runtime and memory measurements, 
we conducted three runs and presented the average across these runs.

A direct comparison to \textsc{SparseDroid} \cite{sparsedroid} was not possible for many reasons. It extends an existing taint analysis client (\myfd{} \cite{flowdroid}) that has a basic integrated alias analysis, whereas our analysis client utilizes a sophisticated external demand-driven pointer analysis \cite{boomerang}. Moreover, \textsc{SparseDroid}'s implementation is not publicly available, and most importantly, IFDS may not terminate when the value domain is infinitely broad.

\section{Related Work} \label{sec:related}
The IFDS \cite{ifds} and IDE \cite{ide} frameworks enabled precise interprocedural data-flow analyses that are flow- and context-sensitive.
Previous works have extended these frameworks with diverse goals. 
Naeem et al. \cite{extendIFDS} proposed four extensions to the IFDS framework, 
to improve its scalability and precision under certain practical analysis conditions.
\myh{} \cite{heros} introduced a Java-based generic IFDS and IDE solver.
\textsc{Reviser} \cite{reviser} proposed an algorithm to adapt IFDS and IDE to incremental program updates.
\textsc{CleanDroid} \cite{cleandroid} introduced a technique for reducing the memory footprint of IFDS-based data-flow analyses.
\textsc{DiskDroid} \cite{diskdroid} applied a disk-assisted computing approach for improving the scalability of IFDS-based taint analysis.

Sparsification has been applied to improve the scalability of static analyses.
Choi et al. \cite{choi1991automatic} introduced sparse data-flow evaluation graphs based on SSA (static-single assignment). 
Oh et al. \cite{sparse-global} presented an abstract interpretation-based framework for designing generic sparse analyses,
which guarantees to preserve the precision of the non-sparse analysis through data dependencies.
\textsc{Pinpoint} \cite{pinpoint}, \textsc{SVF} \cite{svf} and \textsc{SFS} \cite{hardekopf2011millions} utilize cheaper pre-analyses to sparsify pointer analyses. 
Recent on-demand sparsification approaches exploit the data-flow facts that become available during the analysis runtime
for further sparsification. \mysb{} \cite{sparseBoomerang} exploits the variables in alias queries during demand-driven pointer analysis, to create query-specific sparse CFGs. 
The sparse IFDS algorithm \cite{sparsedroid} exploits data-flow facts to create fact-specific sparse CFGs and propagate each fact on its own sparse CFG. In this work, we present the more generic Sparse IDE algorithm that
efficiently solves not just IFDS-based reachability problems, but also IDE problems that require value computation.

\section{Conclusion and Future Work} \label{sec:conc}
In this work, we presented the Sparse IDE framework as
a scalable alternative to the original IDE framework.
Sparse IDE is the first fact-specific sparsification approach that allows for computations on infinitely broad domains.
The essence of Sparse IDE is creating symbol-specific
sparse control flow graphs on-demand, and propagating
data-flow facts sparsely through these graphs. 
Sparse IDE produces equally precise results as the original IDE, 
while significantly improving its scalability.
We also explicitly discuss the limits of sparsification for IDE: while symbol-specific sparsification is possible and useful, one cannot sparsify with respect to the (typically numeric and infinite) value domain.

In the future, we plan to apply the Sparse IDE framework 
to other data-flow analysis problems 
and investigate problem-specific requirements for building sparse CFGs.
We also plan to combine Sparse IDE with other scalability-improving techniques 
that are orthogonal to our sparsification approach.

\begin{acks}
    We gratefully acknowledge the support of Martin Mory and Marcus Hüwe in this work.
    We thank Martin for the enlightening discussions and for the encouragement to conclude this work. 
    We thank Marcus for sharing his expertise on the formal notation.
\end{acks}

\bibliographystyle{ACM-Reference-Format}
\bibliography{acmart}

\end{document}